# Stepwise correlation of multivariate IoT event data based on first-order Markov chains

Vassilis Papataxiarhis
Dpt. of Informatics and
Telecommunications
National and Kapodistrian
University of Athens
Athens, Greece
vpap@di.uoa.gr

Thomais Vassilopoulou
Dpt. of Informatics and
Telecommunications
National and Kapodistrian
University of Athens
Athens, Greece
sdi1500016@di.uoa.gr

Sofia Kostakonti
Dpt. of Informatics and
Telecommunications
National and Kapodistrian
University of Athens
Athens, Greece
sdi1500080@di.uoa.gr

Stathes Hadjiefthymiades
Dpt. of Informatics and
Telecommunications
National and Kapodistrian
University of Athens
Athens, Greece
shadj@di.uoa.gr

*Abstract* — Correlating events in complex and dynamic IoT environments is a challenging task not only because of the amount of available data that needs to be processed but also due to the call for time efficient data processing. In this paper, we discuss the major steps that should be performed in real- or near real-time event management focusing on event detection and event correlation. We investigate the adoption of a univariate change detection algorithm for real-time event detection and we propose a stepwise event correlation scheme based on a first-order Markov model. The proposed theory is applied on the maritime domain and is validated through extensive experimentation with real sensor streams originating from large-scale sensor networks deployed in a maritime fleet of ships.

*Keywords—event correlation; event forecasting; sensor networks; change detection*

## I. INTRODUCTION

In IoT environments, sensor nodes constitute the basic network components that collect and analyze data about the environment within which they are deployed. Sensing elements capture contextual information to support context-aware applications from diverse domains (e.g., maritime, intelligent transportation, smart farming). Being based on the collected information, an IoT network intends to detect events that take place during the network lifetime and may depict, or even affect, the state of the system. The term 'event' is used to describe an alteration on one or more variables monitored by the system (we call them *context attributes - CA*). In this paper, we follow a stepwise approach to analyze the CA and unveil the transitions between the different system states that govern the operation of the network. Such interconnections are normally hidden under the streams of raw data captured by the system.

Sensor streams in IoT environments mostly arrive as sequences of raw data that provide instant measurements or summaries (e.g., mean, max) regarding observed phenomena that may change over time. Due to its frequency, raw data are usually of limited value even for system experts that intend to reach a higher level of understanding for the internal dynamics of the system (e.g., expected sequences of events in the near future, faulty situations, abnormalities). The sensor streams are handled and analyzed in an online fashion in order to identify times when the probability distribution of the respected time series changes. Each sensor stream is transformed into a binary stream that represents time points of unexpected behaviour of a CA *(event detection)*. Event streams can provide refined information of the data that may be of high importance to the decision makers that need to act proactively in order to sustain the proper behavior of the system. For example, the abnormal increase of the values reported by a smoke detector could essentially be an evidence of a fire. As a next step, we focus on the derivation of a dependency structure that describe time-dependent regulations between the incoming data streams and illustrate possible transitions on the system state *(event correlation)*. The resulted rules are extracted from dependency structures that are constructed in near real-time to represent interdependencies among the measured variables *(event prediction)*. The derived dependencies that result in separate time steps are filtered in order to balance between outdated patterns and event sequences of previous steps that could still be of importance *(event filtering)*. Fig. 1 presents the workflow for the online processing of events originated from sensor networks as it is presented in the following sections. In this paper, we focus on the first two steps of the process; event detection and event correlation of multivariate IoT data.

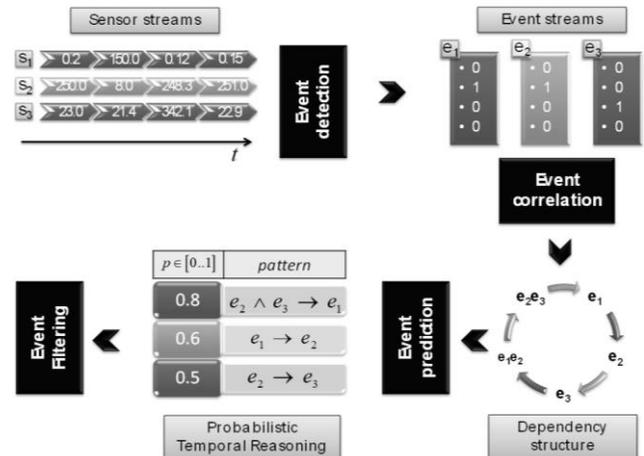

**Fig. 1.** Online event processing in sensor networks.

The contribution of this study is multifold. First, we take advantage of statistical tools to demonstrate that the determination of events in IoT environments can be accommodated within a change detection framework in the context of multivariate time series. As a next step we propose an online event correlation scheme by essentially implementing a first-order Markov chain for identifying the stepwise dependencies of multivariate event data. We investigate the time and space complexity of the algorithm demonstrating its applicability in a real-time monitoring operation of an IoT network. Finally, we evaluate the proposed framework against a variable-order Markov model based on the idea of partial matching [1].

## II. BACKGROUND AND MOTIVATION

Over the last decade, a lot of effort has been devoted in the correlation of event data originated from IoT infrastructures in order to extract patterns that depict the internal dynamics of systems and act proactively. Typical event correlation schemes that operate over univariate time series perform under the following assumptions:

- A transition from object (i.e., event or sequence of events) $A$ to object $B$ occurs if and only if $B$ occurs immediately after $A$ (i.e., not within a time window).
- Only one object is considered at each step of the sequence (i.e., there are no objects occurring at the same time).

However, things may differ in case of event correlation over multivariate sensor data. First, there is no guarantee that events would happen one at a time. On the contrary, an alerting situation or a malfunctioning system is expected to lead to several events triggered at the same time step. For example, consider the case of a fire that is monitored through multiple smoke detectors, temperature and humidity sensors. Most of the sensors that monitor an affected area are expected to experience a significant drift in the probability distribution of their reported values, thus leading to multiple events within the same time steps. Further to the above, since events occur rarely when no alerting situation takes place, the probability of having no events appearing at specific time steps should be also considered.

In [20], we addressed the above challenges by representing the occurrence of event vector streams as a stochastic process. We proposed a *variable-order correlation scheme* based on the idea of *partial matching*, which had initially been applied for web prefetching. The original partial matching algorithm [1] faced the problem of predicting the upcoming $l$ URL accesses of a client based on information about the past $m$ steps. The algorithm maintains a data structure that keeps track of the past URL sequences of length up to $m+l$. The algorithm determines all subsequences in the history structure that match any suffix of the last $m$ accesses. A cut-off probability threshold $p_{thr}$ is used to discard URLs with low probability. All items that match these suffixes exceeding $p_{thr}$ are prefetched. In [20], we implemented and validated a variable-order event correlation algorithm by adapting the idea of partial matching to the context of multivariate streaming sensor data. Our algorithm was based on a variable-order Markov model where Markov chains of orders $1,...,m$ are combined to predict event sequences of length up to $m$. We considered the following parameters:

(a) Number of past multivariate event vectors considered ($m$). It represents the *maximum order* of the model.
(b) Number of steps that the algorithm would predict in the future ($l$).

Based on the results presented in [20], there are a number of cases where the predictability of a time series was decreased when a long history of previous timesteps was considered. In particular, the validation of the partial matching algorithm demonstrated that short-memory processes (i.e., runs with low $m$ values) outperformed the predictions resulted by processes that considered a long history of previous event patterns (high $m$ values). This practically means that statistical processes that are stateful, more complex and take into account a longer history of events do not necessarily lead to better predictions of future system states. Intuitively, this could be explained by considering time series with repeating behavior that result in similar event patterns over time.

In the next chapter we try to address the above challenges in the context of multivariate streaming event data by investigating the problem of event detection and correlation in a stepwise fashion. We consider the occurrence of event vector streams as a stochastic process and we discuss a stateless event correlation scheme based on a first-order Markov chain that represents the subsequent correlation of occurred events.

## III. PRIOR WORK

### A. Statistical process control

Statistical process control constitutes a well-known methodology that has been applied in a wide range of fields to identify detection of changes over numerical data. The purpose of control charts is to allow for the detection of events that represent an actual process change. In general, deciding about the occurrence of a change can be significantly difficult since the process characteristic may be continuously varying. Control charts adapt to the overall process behavior over time and provide statistically objective criteria of change. Several change detection algorithms have been proposed over the years that aim to identify abnormal deviations of the current values with respect to the values arrived in previous steps. Change detection algorithms can be distinguished into the following categories:

*Univariate change detection*. The algorithms belonging in this category take into consideration each sensor stream separately and detect possible abnormalities through a sequential time series analysis. The cumulative sum (CUSUM) algorithm [7] attempts to detect a change on the distribution of a time series with respect to a target value at real-time. CUSUM involves the calculation of positive and negative changes in the time series $x_t$ cumulatively over time and it compares these changes to a positive and a negative threshold. Whenever these thresholds are exceeded, a change is reported through the above-detection and below-detection signals while the cumulative sums are set to zero. In order to avoid the detection of non-abrupt changes or slow drifts, the algorithm takes into consideration tolerance parameters for positive and negative changes. The algorithm assumes that the arrived time series follow a normal distribution, a fact that is not always valid in real and dynamic IoT environments. Similarly to CUSUM, the Shewhart control charts [16] provide a statistical measure to detect abrupt shifts of a time series. In the Shewhart control chart, a variable $x_t \in \mathbb{R}$ is detected to deviate at time $t$ from its normality whenever exceeds one of the control limits specified by the algorithm: the Upper Control Limit (UCL) and the Lower Control Limit (LCL). The control limits are defined as the distance from the current mean value of the statistical process $x_t \in \mathbb{R}$. Shewhart control charts algorithm is presented in Section IV.

*Multivariate change detection.* The algorithms of this category exploit multivariate autoregressive (MAR) models to represent each context vector as a linear sum of previous activity. In particular, the consecutive measurements of a time series contain information about the process that generated it. An attempt for describing this underlying order can be achieved by modelling the current value of the variable $x_t$ as a weighted linear sum of its previous values (i.e., $x_{t-1}, x_{t-2}$, etc.). This is an autoregressive (AR) process and is a very simple, yet effective, approach to time series characterization [21]. The order of the model is the number of preceding observations used in order to determine $x_t$. The weights are the parameters of the model estimated from the data that uniquely characterize the time series. Multivariate autoregressive models (MAR) extend this approach to multiple time series to the point that the vector $\mathbf{x}_t = (x_{1,t}, x_{2,t}, ..., x_{n,t})$ of current values of all variables is modelled as a linear sum of previous vectors. The MAR model describes the dynamic behaviour of time series for forecasting [21]. Then, obtaining a binary value that indicates the change (or non-change) for a specific variable (i.e., sensor stream) is reduced into a single thresholding operation between the estimated future vector and the actual vector arrived.

B. *Event correlation in IoT architectures*

The enormous volume of data produced by modern IoT environments has recently increased the need for intelligent processing and analysis of this data. In particular, extracting meaningful information about collected data and events has gained popularity among both the research and the industrial community the last decade. From a research perspective, literature provides a wide range of approaches for the detection and correlation of events in real-time data management systems. In [22], the author focuses on change detection criteria in multidimensional streaming data. The paper investigates well-known metrics for change detection, such as Kullback-Leibler distance [23] and Hotelling's T-square test for equal means [24], providing a log-likelihood justification. It also proposes and evaluates a semi-parametric log-likelihood criterion that perform better in multidimensional data. In [2], the authors propose a real-time probabilistic framework for the detection of abnormalities in streaming event logs that come from multiple event sources. The proposed scheme assumes the events already detected and the identification of abnormalities is based on the derivation of a directed probabilistic graph that represents the historical footprint of the event occurrences. The graph vertices (i.e., event types) are specified by the prior probabilities for the occurrence of event types while each directed edge comes with a conditional probability between the connected event types. The authors also propose a metric to quantify the similarity between two event streams. A change is detected whenever this metric exceeds a threshold comparing the graph resulted from the recent event records (i.e., stream tail) to the complete graph of the original stream. A similar graph structure is presented in [3] with the edges corresponding to the joint probabilities of two event types (i.e., bidirectional graph). The authors apply measures originated in information theory in order to provide a historical analysis of abnormality events (i.e., events are considered known and detected from historical logs). They also provide metrics and a stepwise analysis for the determination of root-cause events as well as events that can, possibly, lead to a system crash. Both frameworks do not take into account sequences of correlated events since the derived conditional and joint probabilities, respectively, refer to events occurred during the same time window, but not in subsequent time steps. For example, consider a case of an event stream where whenever an event of type A occurs at time *t*, an event of type B also occurs at time step *t+1*. Similarly, implications imposed by joint occurrences of events are not considered due to time complexity (e.g., event types A and B have to occur both in order to result in the occurrence of event type C). In [4], the authors propose a probabilistic scheme for the correlation of distributed events in the network security field. Specifically, they apply a HMM [5] and Kalman filtering [6] to unveil spatial and temporal correlations among the observations and the hidden states of internet security attacks. Kalman fitering and Cumulative Sum algorithm [7] are adopted to detect changes in real time in [8]. The paper also evaluates several criteria for change detection in streaming data.

Industrial maintenance has also received significant attention in academic literature for many decades. Specifically, a lot of studies have been devoted in the area of *condition-based predictive maintenance* [9] for the prediction of faults and errors in complex technical systems including sensor networks. Most of industrial approaches are based on existing statistical methods to identify alerting situations due to abrupt changes of the probability distributions of the observed parameters [10]. In [11], the authors discuss the problem of rare events prediction in multidimensional data and they propose two approaches that address the problem. The first approach focuses on degradation detection (i.e., abnormal system behavior) that is based on one-class Support Vector Machines model [12]. The proposed method formulates a quadratic program that detects anomalies by minimizing the distance of each new multidimensional data vector with regard to a training dataset. Then, a moving average model is used for modeling multidimensional degradation behavior in time. The second approach exploits a regularized logistic regression classification method where the predictor function consists of a transformed linear combination of explanatory variables. Both approaches are evaluated over datasets resulted by real-world cases from aircraft operational performance area. In [13], the authors propose one model-based and one data driven method for predicting faults in multi-sensor systems. The former approach is based on an autoregressive moving average model where the estimated model parameters are further used for implementing the change detector that is realized as a Neyman-Pearson hypothesis test [14]. The second approach takes advantage of a HMM and Viterbi algorithm [15] to estimate the most likely sequence of hidden states. Both approaches are validated through simulations with synthetic data.

## IV. METHODOLOGY

### A. Event detection based on context changes

We observe in real-time, with some uniform frequency, multivariate time series of quantitative system performance parameters. Let $s_i$ be a *sensor stream* reporting numeric sensor values and let $s_i(t)$ be the value of stream $s_i$ at time $t$, $t \geq 0$. Assuming that $n$ sensor streams are synchronized to report their values periodically, we represent the set of multidimensional contextual information at each time $t$ by the *context vector (CV)*, $CV_t = (s_1(t), s_2(t), ..., s_n(t)) \in \Re^n$. Practically, each sensor stream formulates a univariate time series while the CV stream represents a multivariate time series. Event detection aims to determine the values $s_i(t)$ that constitute abrupt changes within a CV stream. Specifically, we transform each CV of length $n$ into a binary vector of the same length with each value representing a possible change in the respective stream. We call such deviations of the normal behavior *events* and the binary vector *event vector (EV)*. Let $I = \{e_1, ..., e_n\}$ be the set of the considered event types $e_i$ with regard to the respective sensor streams $s_i$, $i \in [1, n]$. An event can be any observation that does not conform to an expected pattern. An event vector at time $t$ is represented by $EV_t = (e_1^t, e_2^t, ..., e_n^t) \in \{0,1\}^n$, where $e_i^t = e_i(t)$ is the binary value representing whether an abnormal behavior took place for stream $s_i$ (value '1') or not (value '0').

The transformation of a CV into an EV is based on change detection algorithms that aim to identify abnormal deviations of the current values with respect to the values arrived in previous steps In the context of this work, we have experimented with Shewhart control charts [16] that considers each sensor stream separately (*univariate change detection*). Shewhart control charts [16] provide a statistical measure to detect abrupt shifts of a time series. In the Shewhart control chart, a variable $x_t \in \mathbb{R}$ is detected to deviate at time $t$ from its normality whenever exceeds one of the control limits specified by the algorithm: the Upper Control Limit (UCL) and the Lower Control Limit (LCL). The control limits are defined as the distance from the current mean value of the statistical process $x_t \in \mathbb{R}$. Specifically, UCL and LCL are defined as follows:

$$UCL = \bar{x}_t + k \cdot \sigma_t$$
$$LCL = \bar{x}_t - k \cdot \sigma_t \quad (1)$$

where $\bar{x}_t$ represents the average of the time series at time $t$ and $\sigma_t$ is the standard deviation at the same time step. The parameter $k$ represents the tightness of the change detection process. Low values of $k$ lead to tight control of the measurement process while large values flag only those measurements that are egregiously out of control.,

At a time step $t$, the data stream $x_t$ is detected to trigger an alarm if $x_t > UCL$ or $x_t < LCL$. The controller returns an output detection signal $s$ at each step $t$ where $s=1$ if there is a change detected; in case of normality (i.e., $x_t \in [LCL, UCL]$), the algorithm sets $s=1$. Shewhart control chart is presented in Algorithm 1 below.

---
**ALGORITHM 1.** Shewhart Control Chart
**Input:** univariate time series $x_t$, tightness $k$
**Output:** detection signal $s$

1: $\bar{x}_0 \leftarrow 0$;
2: $\sigma_0 \leftarrow 0$;
3: $t \leftarrow 1$;
4: **while** ( *true* )
5: $\quad \bar{x}_t \leftarrow \bar{x}_{t-1} + \dfrac{x_t - \bar{x}_{t-1}}{t}$;
6: $\quad \sigma_t \leftarrow \sqrt{\dfrac{1}{t}\left((t-1) \cdot \sigma_{t-1}^2 + (x_t - \bar{x}_t)(x_t - \bar{x}_{t-1})\right)}$;
7: $\quad UCL_t \leftarrow \bar{x}_t + k \cdot \sigma_t$;
8: $\quad LCL_t \leftarrow \bar{x}_t - k \cdot \sigma_t$;
9: $\quad$ **if** (($x_t > UCL$) **or** ($x_t < LCL$)) **then**
10: $\quad\quad s \leftarrow 1$;
11: $\quad$ **else**
12: $\quad\quad s \leftarrow 0$;
13: $\quad$ **end**
14: $\quad t \leftarrow t+1$;
15: **end**

---

Fig. 2 illustrates an example of the Shewhart controller over a sensor stream $x_t \in \mathbb{R}$. Shewhart control chart has to be applied to each variable separately in case of multivariate time series. However, the algorithm does not assume normal distribution for $x_t$. This makes the algorithm quite robust in real-life datasets where most of the times there is no knowledge available for the probability distribution that a data stream follows.

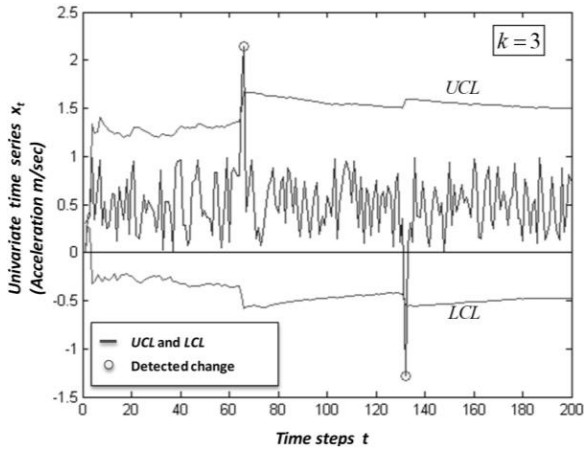

**Fig. 2.** Shewhart change detection chart.

### B. Stepwise event correlation and forecasting

Taking into account the requirements imposed by multivariate time series, we propose the use of Markov chains to correlate event data in a stepwise fashion. The stepwise correlation of events is essentially realized through a first-order Markov chain that represents event patterns that occurred in the past in order to predict future sequences of events.

Let $I = \{e_1, ..., e_n\}$ be the set of the considered event types $e_i$ with regard to the respective sensor streams $s_i$, $i \in [1, n]$. We represent the $n$–dimensional event vector at time $t$ by $EV_t = (e_1^t, e_2^t, ..., e_n^t) \in \{0,1\}^n$, where $e_i^t = e_i(t)$ are the binary values resulted from the change detection step and they represent whether an abnormal behavior took place for stream $s_i$ (value '1') at time $t$ or the value $s_i^t = s_i(t)$ that arrived was in the expected range. Considering the occurrence of a specific set of events at a certain time step, we can represent the prior probabilities of each subset $I_s \subseteq I$ as the frequency $N(I_s | \overline{1:t})$ of the joint occurrences of all the event types $e_i \in I_s$ and the non-occurrences of all the rest types $e_j \in I - I_s$ from the beginning of the event trace divided by the total number of time steps so far.

$$(\forall t \in N)\ (\forall I_s \subseteq I)\ \ P_{I_s}^{\overline{1:t}} = P(I_s, \overline{1:t}) =$$
$$= \frac{N(I_s | \overline{1:t})}{t} = \frac{\sum_{k=1}^{t} \prod_{e_i \in I_s, e_j \in \bar{I}_s} e_i^k \cdot \bar{e}_j^k}{t} \quad (2)$$

In the above form, the binary value $\bar{e}_j^k = 1 - e_j^k$ represents the boolean negation of the binary value $e_j^k$, while $\bar{I}_s = I - I_s$. Obviously, Eq.(2) quantifies the probability of joint occurrences for exactly the event types that belong to a specific event set across the event stream.

Similarly, we can represent the conditional probabilities between two subsets $I_l, I_m \subseteq I$ as the frequency $N(I_l^{t-1}, I_m^t | \overline{1:t})$ of the joint occurrences of both event sets in subsequent time steps divided by the total number of occurrences of the event set $I_l$ until time step $t-1$.

$$(\forall t \in N)(\forall I_l, I_m \subseteq I)\ P_{I_l I_m}^{\overline{1:t}} = P(I_m | I_l, \overline{1:t}) =$$
$$\frac{N(I_l^{t-1}, I_m^t | \overline{1:t})}{N(I_l | \overline{1:t-1})} = \frac{\sum_{k=2}^{t} \prod_{e_i \in I_l, e_j \in \bar{I}_l} \prod_{e_p \in I_m, e_q \in \bar{I}_m} e_i^{k-1} \cdot \bar{e}_j^{k-1} \cdot e_p^k \cdot \bar{e}_q^k}{\sum_{k=1}^{t-1} \prod_{e_i \in I_l, e_j \in \bar{I}_l} e_i^k \cdot \bar{e}_j^k} \quad (3)$$

The above definitions of probability measures are the basic tools for the quantification of event sequences and correlation among events. Specifically, we model the probabilistic behavior of event occurrences by a time discrete stochastic process as presented below.

Let $X(t)$ be a random variable representing the state of the system at each time step $t$, i.e., $(t \in N)\ X(t) \in \{0,1\}^n$. The possible values of $X(t)$ form the countable state space of the chain. We represent the state space by $S = \{0,1\}^n$. We also assume the discrete stochastic process $\{X(t), t \in N\}$ equipped with the Markov property (*memoryless*), i.e., each future state depends only on the current state:

$$(\forall t \in N)\ P(X(t+1) = EV_{t+1} | X(t) = EV_t, ..., X(1) = EV_1) =$$
$$P(X(t+1) = EV_{t+1} | X(t) = EV_t) \quad (4)$$

The above stochastic process can be, intuitively, depicted by a directed graph $G = \langle V, E \rangle$ where $V = \mathbf{P}(I)$ is the powerset of $I = \{e_1, ..., e_n\}$ that represents the states of the process and $E$ is a set of directed edges among the states representing stepwise correlation (i.e., occurrence in the subsequent time step).

Each *state vertex* $v \in V$ can be represented as a tuple of the form:

$$v = \langle I_v, P_v^t \rangle,\ I_v \subseteq I,\ P_v^t = P_{I_v}^{\overline{1:t}} \quad (5)$$

where $I_v$ is a specific set of events that take place in this state of the process and $P_v^t$ is the prior probability of $I_v$ as defined in Eq.(2) for the time step $t$. Similarly, we represent each *weighted transition edge* $d_{uv} \in E$ as follows:

$$(\forall u, v \in V)\ d_{uv} = \langle u, v, P_{uv}^t \rangle,\ I_u, I_v \subseteq I,\ P_{uv}^t = P_{I_u I_v}^{\overline{1:t}} \quad (6)$$

where $I_u, I_v$ represent event sets that occur in subsequent time steps and $P_{uv}^t$ is the conditional probability between those sets for the time step $t$ as it was defined in Eq.(3).

The above probabilistic graph model depicts the stepwise correlation between the states of the process in subsequent time steps. As a Markov chain, the set of graph nodes represents a partition of the probability sample space with each node corresponding to a singleton unit set. As a result, the prior probabilities of all states sum up to 1 in every step.

$$(\forall t \in N)\ \sum_{v \in V} P_v^t = 1 \quad (7)$$

The same stands for the probabilities of the outgoing edges for each node of the graph.

$$(\forall t \in N)(\forall v \in V) \sum_{u \in V} P_{vu}^t = 1 \quad (8)$$

We should highlight here that the above frequencies can be calculated in a cumulative manner with regard to the expansion of the event streams, thus the graph can be updated on the arrival of new event vectors.

A walk within the graph starting by the current state represents probability of future event sequences. Specifically, given the prior and conditional probabilities of the current step $t$, the next step prior probabilities can be estimated as follows:

$$(\forall t \in N)(\forall v \in V) \quad P_v^{t+1} = \sum_{k \in V} P_k^t \cdot P_{kv}^t \quad (9)$$

Additionally, we can estimate the probability of occurrence for a specific set of event types $I_s$ (i.e., not a state) at the next step by the following form:

$$(\forall t \in N)(\forall I_s \subseteq I) \quad P_{I_s}^{t+1} = \sum_{I_v \supseteq I_s} \sum_{k \in V} P_k^t \cdot P_{kv}^t \quad (10)$$

Obviously, the considered Markov chain is not time-homogeneous since the transition probabilities change over time. However, we may take advantage of the Chapman-Kolmogorov equations to provide estimations for the $n$-step conditional probabilities, as follows:

$$(\forall n, m \in N)(\forall u, v \subseteq V) \quad P_{uv}^{n+m} = \sum_{k \in V} P_{uk}^n \cdot P_{kv}^m \quad (11)$$

Figures 3 and 4 illustrate an example of the stepwise event correlation algorithm. The event vectors refer to three types of events (A, B and C) that they arrive in the order illustrated in the respective tables included in the figures. The graph nodes represent states either containing single events or combination of events since at each step more than one event may occur. For the sake of simplicity, graph nodes and edges with prior and conditional probabilities, respectively, that equal to zero are omitted.

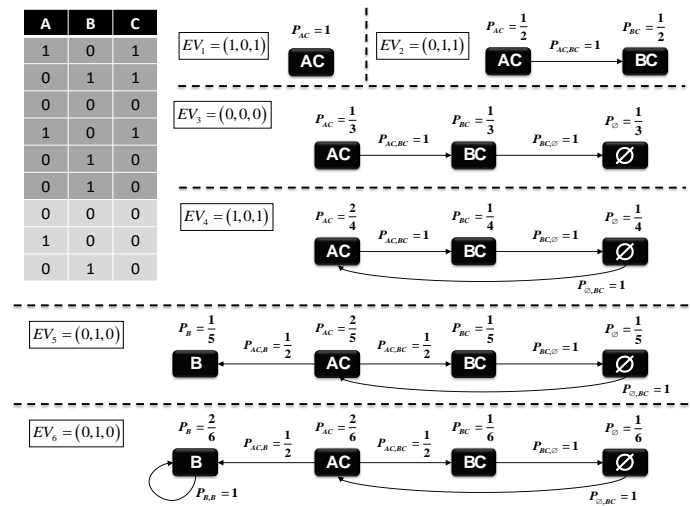

**Fig. 3.** Example of stepwise correlation algorithm for multivariate event data – Steps 1-6.

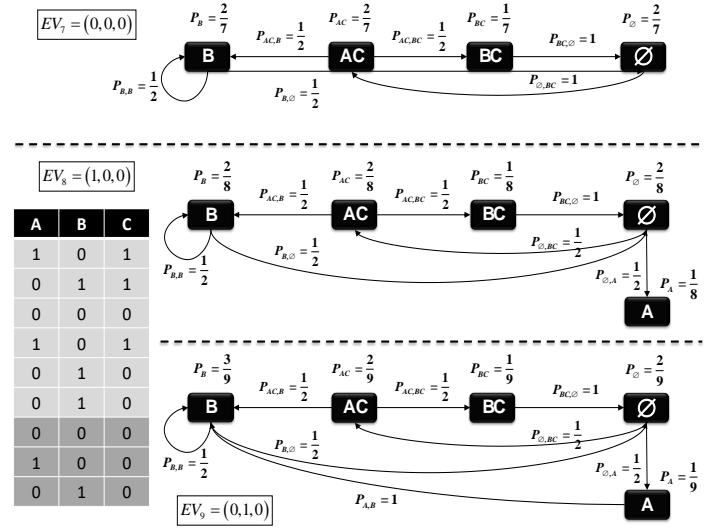

**Fig. 4.** Example of stepwise correlation algorithm for multivariate event data – Steps 7-9.

The stepwise correlation algorithm assumes, in the worst case, that all prior probabilities (i.e., all $|V|$ nodes) and conditional probabilities originated from the previous state (i.e., $|V|$ edges at most) should be examined at each step. Since $|I| = n$, the size of set $V$ becomes exponential in the case of a fully connected graph. Specifically, $|V| = \sum_{i=0}^{n} \binom{n}{i} = 2^n$. However, storing all possible states is of little use in practice since the probability of occurrence of a big number of events at the same time step is practically very low. If we consider only combinations consisting of $k$ events at most (with $k$ fixed and independent of $n$), then the upper limit on the number of the considered nodes is significantly reduced to the following estimation:

$$|V| = \sum_{i=0}^{k} \binom{n}{i} = \binom{n}{0} + \binom{n}{1} + \binom{n}{2} \ldots + \binom{n}{k} =$$
$$= 1 + n + \frac{n \cdot (n-1)}{2} + \ldots + \frac{\prod_{j=0}^{k-1}(n-j)}{k!(n-k)!} = O(n^k) \quad (12)$$

This means that, assuming that combinations of $k$ event types at most (with $k$ fixed) will occur, the time complexity of the algorithm is polynomial. Since the event frequencies can be updated cumulatively in an online fashion, the algorithm is able to construct its history structure on the fly by continuously updating it for every new event vector arrived.

In this section we presented a graph structure equivalent to a first-order Markov chain representation for correlating multivariate sensor data. As a result, the prediction of future events is based only on the current state of the process without taking into account preceding states. The next section presents a variable-order algorithm that intends to address this issue.

## V. EXPERIMENTAL EVALUATION

The proposed event correlation scheme was evaluated in a number of different real-world scenarios using real data coming from the maritime domain. Specifically, our scheme has been assessed in real-world scenarios where the machinery and the infrastructure of three (moving) ships were constantly monitored making this information available for further analysis and processing at the back-end in real-time. At each timestep, 29 environmental parameters were measured. The 29 sensor streams provided more than 3.5MB of data per minute resulting in a total of more than 450GB for a period of 3 months (for all the three ships). The considered 29 parameters included ship speed, acceleration, direction, wind speed, machine temperature and humidity, person count, gas percentage, GPS, depth levels, and water inclination.

Since the data was not following the normal distribution, we decided to use Shewhart control charts to detect potential abnormalities in the sensor streams. For each numeric stream, the result of the change detection process was a binary stream that was provided as input to the event correlation algorithm. We performed several tests to measure the *precision* and the *recall* of the proposed first-order Markov chain following the definitions below:

$$Precision = \frac{TP}{TP+FP}, \text{ and } Recall = \frac{TP}{TP+FN} \quad (13)$$

where *TP* refers to the correctly indicated predictions of the future system state (*true positives*), *FP* refers to the states that were falsely predicted as future states (*false positives*) and *FN* represents the future system states that were not predicted by the algorithm (*false negatives*).

Each precision and recall value was measured in 21144 sequential time steps that equals to the number of available numerical context vectors (i.e., the rows in the database). At each step, the prediction recommended by the algorithm was the one with the highest probability value as it was indicated by the event correlation scheme. Finally, the predictions for the current time step have been checked of whether they became valid after the considered period of time. Based on these results, the precision and recall values were updated accordingly in each step and the average values for each experiment are presented in the following figures.

We also assessed the quality of future event predictions based on different values of the fixed combinations parameter (i.e., *k*) that was considered each time. Figures 5 and 6 demonstrate the performance of the proposed forecasting scheme over different time horizons (*h*). Similarly to the *l* parameter of the partial matching algorithm, we define time horizon *h* as the number of steps within which we validate whether an event took place or not. For instance, when *h*=1 predictions made at timestep *t* are expected to take place in the next timestep *t+1*. On the other hand, when *h*=3 predictions made at timestep *t* are expected to take place within the next three timesteps between [*t+1, t+3*]. Obviously, as *h* increases we expect both the precision and recall values to also increase since we make our predictions more reluctant over time. It is worth mentioning that only exact matches of system states are considered as successful predictions. This practically means that if, for instance, *k*=3 and stepwise correlation algorithm predicted that the next system state will consist of two specific types of events (let's assume *event_type_A* and *event_type_B*) then only the occurrence of the exact two types of events is considered as a successful prediction. If not all the event types occur (e.g., only *event_type_B*) or more event types take place (e.g., *event_type_C* as well) in the considered time horizon then we consider it as a false prediction.

Figure 5 and 6 present the results of precision and recall, respectively, over a varying number of considered events at the same timestep. For instance, in the case of Fig. 5, the precision value ranges between 18.2% and 61.9% for *h*=1while the respective curve for recall in Fig. 6 ranges between 14.4% and 51.1%. A first and clear result that can be deduced from the figures is that the algorithm performed better for *low* and *high* values of *k*. Low values of *k* practically refer to system states represented by isolated events that did not affect the rest of the monitored parameters. Such isolated events happened rarely and the algorithm was capable of predicting such events with quite good results both in terms of precision and recall. The same stands for system states consisting of a high number of events ( $k \geq 5$, rare system states with clear correlation between the monitored variables). On the other hand, stepwise correlation did not perform well for intermediate values of *k* (*k*=2,3,4). This is explained by the fact that more than 2 events happened in the vast majority of events. If *k* value was not selected high enough to capture all of the events then the algorithm had to randomly choose between all the events that took place. This normally led to false predictions in the next timesteps since several types of events were not taken into account. Hence, out first clear result is the fact that *k* value should be chosen sufficiently high to consider all the potential events that could take place at the same timestep.

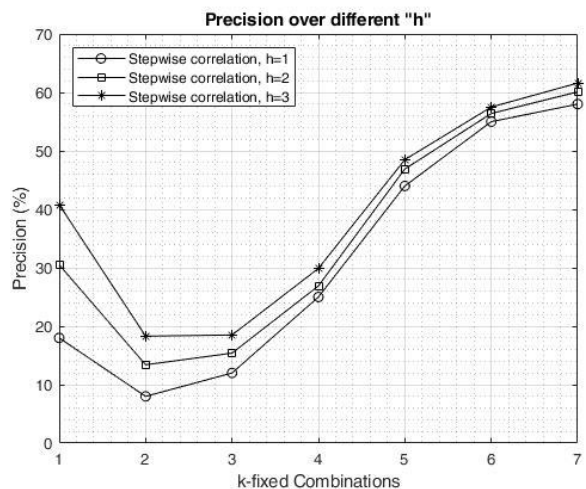

**Fig. 5.** Precision validated over different time horizons *(h)*.

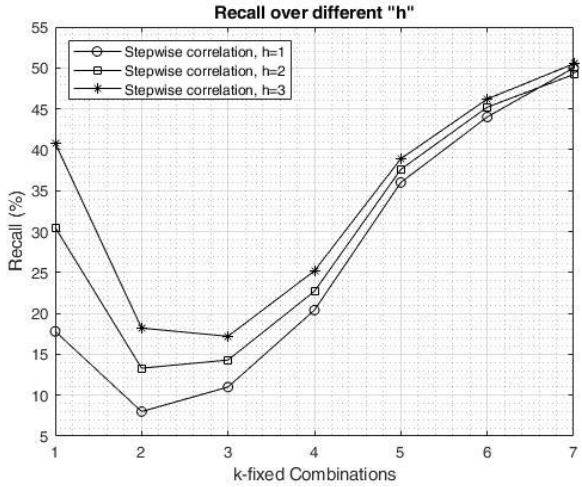

**Fig. 6.** Recall validated over different time horizons *(h)*.

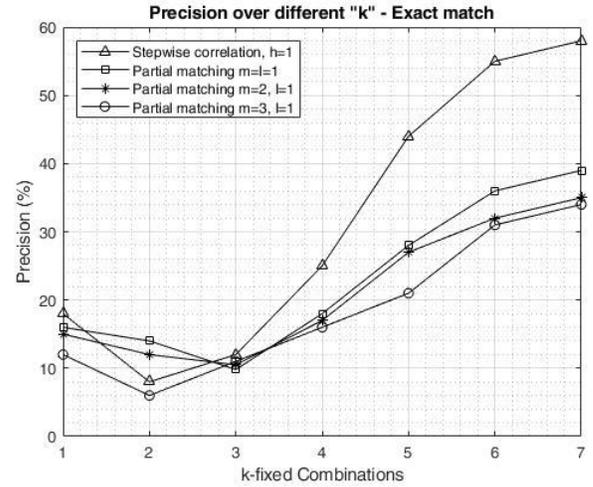

**Fig. 7.** Precision validated with fixed time horizon *(h=1)* varying number of considered events *(k)*.

In the next series of experiments we validated the stepwise correlation algorithm against the event correlation based on partial matching. We experimented with both algorithms over different *k* values (k-fixed combinations) considering fixed time horizons (*h* and *l*, respectively). We also experimented with the memory of the partial matching algorithm (*m* value) to check whether considering longer traces of events can lead to better predictions of the future system states.

Fig. 7 and 8 present the precision and the recall results for different *m* values considering *h=l=*1 for the two algorithms. Fig. 7 shows that only for *k*=2 the partial matching algorithm performed better than the stepwise correlation scheme in terms of precision. In all other cases, stepwise correlation managed to achieve better results than partial matching. For $k \geq 4$ the stepwise correlation scheme clearly dominated in terms of precision. Fig. 8 shows that for *k*=2,3 the partial matching algorithm performed better than the stepwise correlation scheme in terms of recall. This result can be explained by the fact that the algorithm failed to perform well enough for low *k* values because not all the events happened concurrently could be adequately handled by the stepwise dependency graph. Similarly to precision, the stepwise correlation scheme clearly dominated in terms of recall for $k \geq 5$.

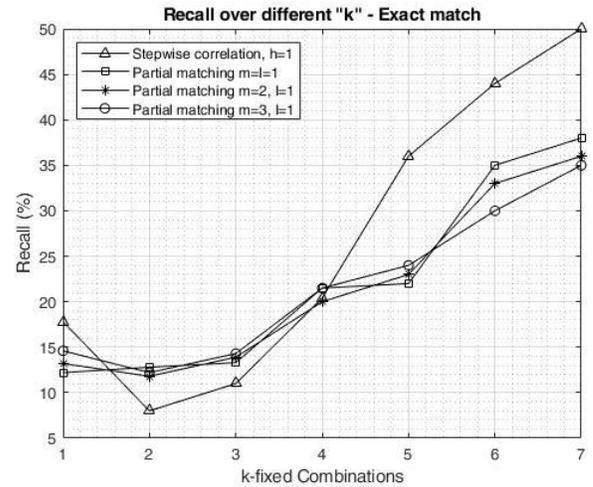

**Fig. 8.** Recall validated with fixed time horizon *(h=1)* varying number of considered events *(k)*.

Fig. 9 and 10 present the precision and the recall results for different *m* values considering same time horizons *h=l=*3 for the two algorithms. Fig. 9 shows that only for *k*=3 the partial matching algorithm performed better than the stepwise correlation scheme in terms of precision. In all other cases, stepwise correlation managed to achieve better results than partial matching. For $k = 1$ and $k \geq 4$ the stepwise correlation scheme clearly dominated in terms of precision. Fig. 10 shows that for *k*=3 the partial matching algorithm performed slightly better than the stepwise correlation scheme in terms of recall. Similarly to Fig. 8, here the result can be explained by the fact that the algorithm failed to perform well enough for low *k* values because the dependency graph did not manage to appropriately represent all the system states. Similarly to precision, the stepwise correlation scheme clearly dominated in terms of recall for $k = 1$ and $k \geq 4$.

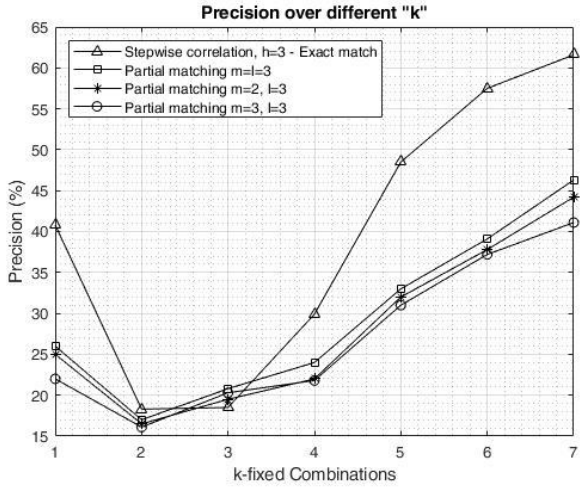

**Fig. 9.** Precision validated with fixed time horizon *(h=3)* varying number of considered events *(k)*.

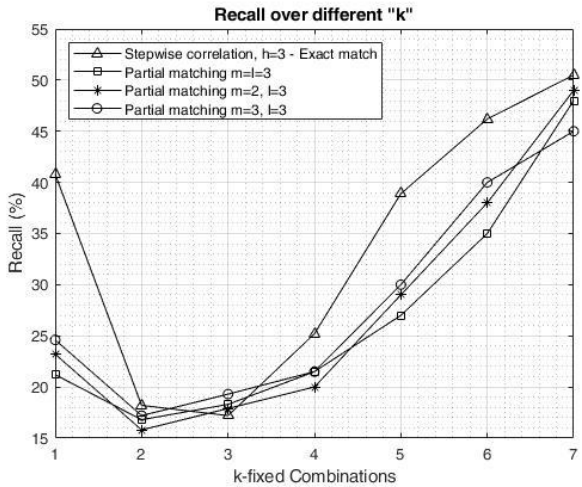

**Fig. 10.** Recall validated with fixed time horizon *(h=3)* varying number of considered events *(k)*.

The last set of experiments we conducted aims to unveil the internal dynamics of the considered dataset by investigating its predictability over time. In particular, the results presented between Figures 7 and 10 showed that even if the stepwise correlation is based on a stateless process (i.e., a first-order Markov model) it seems to outperform higher-order schemes with memory as the one of the partial matching algorithm [1] [20]. In other words, the predictability of the monitored time series was decreased when a long history of previous timesteps was considered. This could be intuitively justified by having a repeating system behavior that, in turn, leads to repeating event patterns that take place over time. The figures below intend to explain the above fact by visualizing three self-similarity metrics for three of the monitored parameters.

The *Hurst exponent* (*H*) is a statistical measure used to classify time series and infer their predictability, i.e. the level of difficulty in predicting future values [25]. The closer the Hurst exponent value is to 0, the stronger is the tendency for the time series to revert to its long-term means value. In other words, an increase will most likely be followed by a decrease and vice-versa. Similarly, a Hurst exponent value close to 1.0 indicates persistent behavior; the larger the *H* value the stronger the trend. In that case, an increase in values will most likely be followed by an increase in the short term and a decrease in values will most likely be followed by another decrease in the short term [26].

On the other hand, autocorrelation [27] and partial autocorrelation metrics [28] constitute statistical measures for estimating the self-similarity of a time series. Autocorrelation measure intends to calculate the correlation for time series observations with observations of previous time steps, called *lags*. In particular, autocorrelation of lag *k* is the correlation between $x_t$ and $x_{t+k}$. On the other hand, the partial autocorrelation of lag *k* estimates the conditional correlation between $x_t$ and $x_{t+k}$ given the values of $x_{t+1}$, $x_{t+2}$, ..., $x_{t+k-1}$. The above statistical measures represent different aspects of self-similarity within a univariate time series.

Fig. 11, 13 and 15 represent samples of three different parameters included in the considered dataset from the maritime domain along with the Hurst exponent values of each time series. The three monitored variables correspond to the depth offset of the ship, the ship speed and the wind speed, respectively. In addition, Figures 12, 14, and 16 present the autocorrelation and partial autocorrelation values for the considered variables. A first and clear result is the fact that in the considered dataset there is a clear relationship between the Hurst exponent and the autocorrelation values. In particular, we can assume that the Hurst exponent, *H*, is <0.5 for rough anti-correlated series, >0.5 for positively correlated series, and ~0.5 for random, white noise time series. For instance, we can observe that the ship speed parameter results in both high *H* values and high autocorrelation (Fig. 15 and 16, respectively). Similarly, the wind speed parameter results in high values for *H* and autocorrelation (Fig. 13 and 14, respectively). On the other hand, the depth offset of the ship with frequent peaks results in lower *H* and autocorrelation value (Fig. 11 and 12, respectively). This relationship explains the tendency of certain parameters within the dataset to have repeating behaviors and show similar patterns over time. It also explains the fact that stateless forecasting schemes such as the stepwise correlation presented in this paper tend to perform better than algorithms based on long history of event traces.

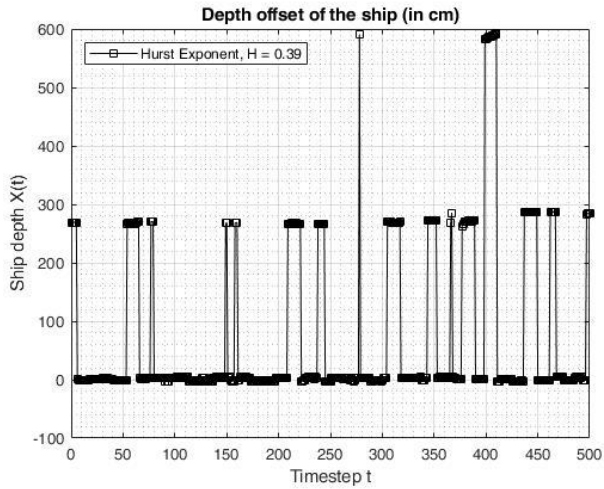

**Fig. 11.** Hurst exponent of the ship depth parameter.

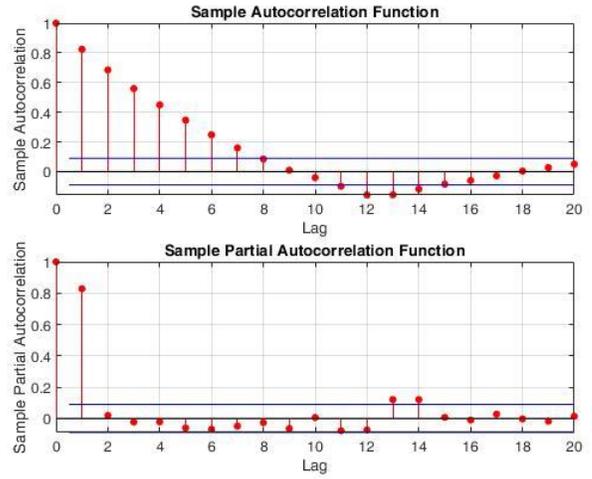

**Fig. 12.** Autocorrelation and partial autocorrelation for ship depth parameter.

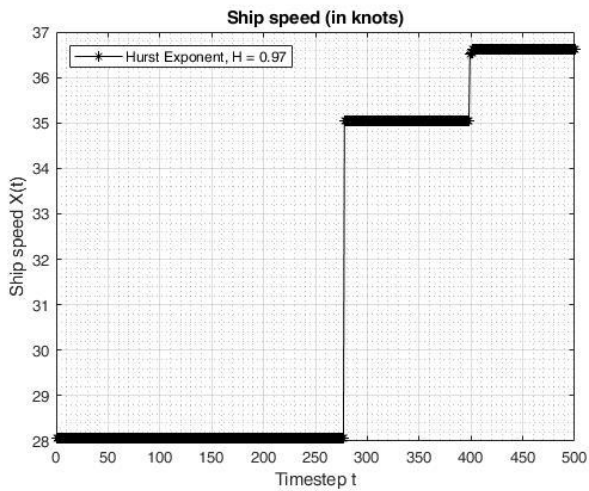

**Fig. 13.** Hurst exponent of the ship speed parameter.

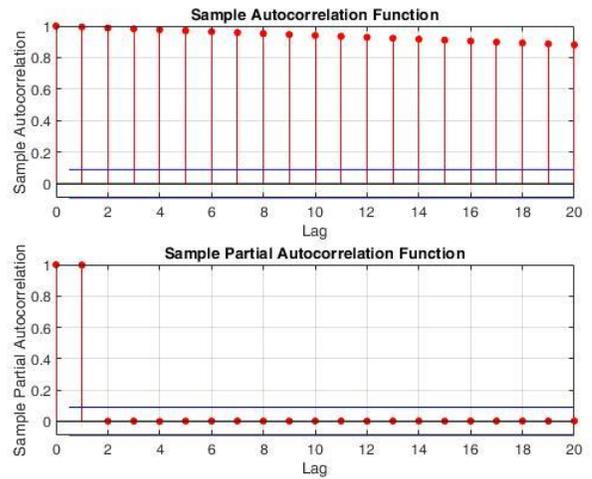

**Fig. 14.** Autocorrelation and partial autocorrelation for ship speed parameter.

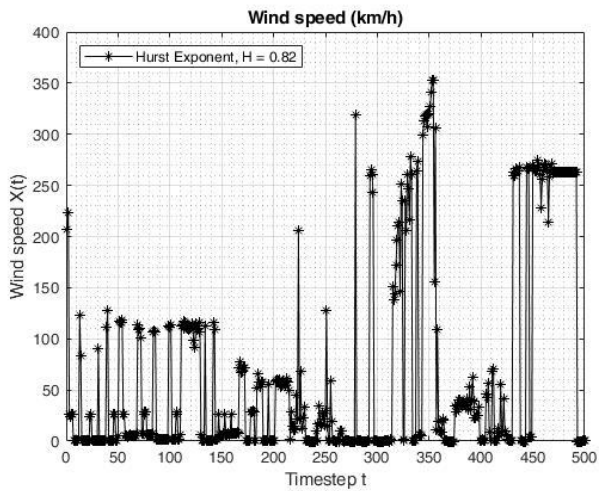

**Fig. 15.** Hurst exponent of the wind speed parameter.

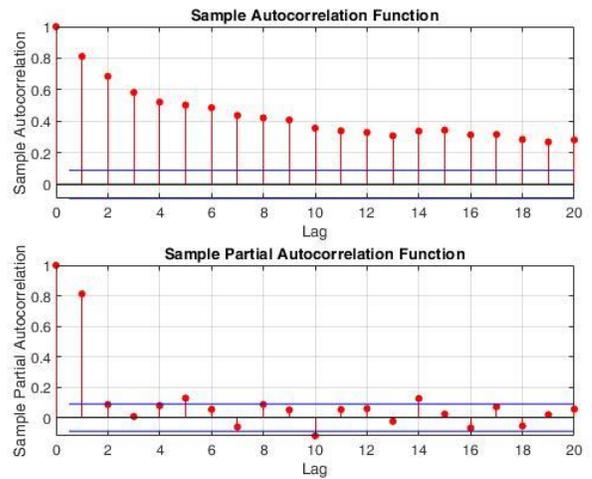

**Fig. 16.** Autocorrelation and partial autocorrelation for wind speed parameter.

## VI. Conclusion

The paper discussed several parts of the processing chain in the context of online event detection and correlation over sensor data in IoT. First, we demonstrated that change detection techniques can be used for the determination of events in sensor networks. Then, we proposed a stateless algorithm for correlating multivariate event data. The algorithm involved a first-order Markov model to capture event sequences. We also discussed the predictability of multivariate time series and we evaluated the proposed algorithm based on real-world data from the maritime domain. The results compared against a variable-order Markov forecasting model.

The presented correlation scheme faced the problem of correlating multivariate event data by, essentially, adopting first-order Markov models to represent frequent event patterns. Since these models capture sequences of events that take place in subsequent steps, they have certain limitations in representing direct dependencies among events that occur within a larger timeframe. In most cases correlated events may not happen one after the other since the effects of the causal may need time to be observed. For example, we can consider the case where an event of type A which is coupled with a temperature sensor occurs exactly ten time steps after the occurrence of event type B related to a smoke detector. In that case, even if a ten-order model could capture this dependency through multiple time steps, it could not provide a direct correlation between the cause (i.e., increase of temperature) and the effect (i.e., existence of smoke). As a next step, we plan to investigate approaches that can address time dependencies among events within larger timeframes. A possible alternative could take advantage of a sliding window that would add memory to the algorithm.

Evidence of a bad or malfunctioning behavior of a sensor network is often embedded within sequences of events detected in distributed nodes across the network. Such state of importance may involve multiple time steps in order to reveal its effects. Also, several bad states could lead to the similar types of events. Our future research agenda will focus on the identification of hidden (i.e., unobservable) system states that may affect the observable variables (i.e., sensors). Hidden Markov Models set up a modeling framework for the representation of such causalities and the determination of unobservable states of a system with regard to the observed measurements.

## References

[1] L. Fan, P. Cao, W. Lin, and Q. Jacobson. 1999. Web prefetching between low-bandwidth clients and proxies: potential and performance, in the *Proceedings of the 1999 ACM SIGMETRICS international conference on Measurement and modeling of computer systems*, Pages 178-187 ACM New York, NY, USA ©1999, DOI: 10.1145/301453.301557.

[2] A.N. Harutyunyan, A.V. Poghosyan, N.M. Grigoryan, and M.A. Marvasti, "Abnormality analysis of streamed log data", in IEEE/IFIP Netwotk Operations and Management Symposium (NOMS), 5-9 May, Krakow, Poland, 2014.

[3] M.A. Marvasti, A.V. Poghosyan, A.N. Harutyunyan, and N.M. Grigoryan, "An anomaly event correlation engine: Identifying root causes, bottlenecks, and black swans in IT environments", VMware Technical Journal, vol. 2, issue 1, pp. 35-45, 2013.

[4] G. Jiang, and G. Cybenko. 2004. Temporal and Spatial Distributed Event Correlation for Network Security, In the Proc. of *2004 American Control Conference*, pages 996 – 1001, Vol. 2 Boston, June 30 - July 3.

[5] O. Cappe, E. Moulines, and T. Ryden. 2005. *Inference in Hidden Markov Models*, 2005:Springer-Verlag.

[6] R. E. Kalman. 1960. A New Approach to Linear Filtering and Prediction Problems. *Journal of Basic Engineering* **82** (1): 35–45. doi:10.1115/1.3662552

[7] E. S. Page. 1954. Continuous Inspection Scheme. Biometrika 41 (1/2): 100–115. doi:10.1093/biomet/41.1-2.100. JSTOR 2333009.

[8] M. Severo and J. Gama. 2006. Change Detection with Kalman Filter and CUSUM, In the *Proc. Int'l Conf. Discovery Science*, pp. 243-254, 2006.

[9] J. Veldman, H. Wortmann, W. Klingenberg. 2011. Typology of condition-based maintenance . *Journal of Quality in Maintenance Engineering* , 17 (2), 183 - 202. 10.1108/13552511111134600.

[10] M. Basseville, and I. V. Nikiforov. 1993. *Detection of Abrupt Changes: Theory and Application*. Prentice-Hall, Inc., Upper Saddle River, NJ, USA (1993).

[11] S. Alestra, C. Bordry, C. Brand, E. Burnaev, P. Erofeev, A. Papanov, and D. C. Silveira-Freixo. 2014. Rare event anticipation and degradation trending for aircraft predictive maintenance, In the *Proceedings of 11th World Congress on Computational Mechanics (WCCM XI)*, July 21, 2014.

[12] I. Steinwart and A. Christmann. 2008. *Support Vector Machines*. Springer, 2008.

[13] A. Marjanovic, G. Kvascev, P. Tadic, Z. Durovic. 2011. Applications of Predictive Maintenance Techniques in Industrial Systems, *Serbian Journal of Electrical Engineering Vol. 8, No. 3*, November 2011, 263-279, DOI: 10.2298/SJEE1103263M.

[14] K. Fukunaga. 1990. *Introduction to Statistical Pattern Recognition*, Academic Press, San Diego, USA, 1990.

[15] A. J. Viterbi .1967. Error bounds for convolutional codes and an asymptotically optimum decoding algorithm. *IEEE Transactions on Information Theory 13 (2): 260–269*. DOI:10.1109/TIT.1967.1054010.

[16] D. Montgomery. 2005. *Introduction to Statistical Quality Control*. Hoboken, New Jersey: John Wiley & Sons, Inc. ISBN 978-0-471-65631-9. OCLC 56729567.

[17] P. Shakarian, A. Parker, G. I. Simari, V. Subramanian, Annotated probabilistic temporal logic. *ACM Transactions on Computational Logic*, 12:14:1–14:44, 2011.

[18] I. Koychev. 2000. Gradual forgetting for adaptation to concept drift. In Proceedings of ECAI Workshop Current Issues in Spatio-Temporal Reasoning, Berlin, Germany, pp. 101{106. ECAI Press.

[19] R. Klinkenberg. 2004. Learning drifting concepts: Example selection vs. example weighting. Intelligent Data Analysis 8 (3), 281-300, 2004.

[20] V. Papataxiarhis, S. Hadjiefthymiades, "Event correlation and forecasting over high-dimensional streaming sensor data", 2018 IEEE 14th International Conference on Wireless and Mobile Computing, Networking and Communications (WiMob 2018), Limassol, Cyprus, October 15-17, 2018.

[21] H. Luetkepohl. 1991. Introduction to multiple time series analysis. Springer, New York, pp 368–370.

[22] L. Kuncheva. 2013. Change Detection in Streaming Multivariate Data Using Likelihood Detectors. *IEEE Transactions on Knowledge and Data Engineering*, Volume 25, Issue 5, pages 1175 – 1180, DOI: 10.1109/TKDE.2011.226.

[23] T. Dasu, S. Krishnan, S. Venkatasubramanian, and K. Yi. 2006. An Information- Theoretic Approach to Detecting Changes in Multi-Dimensional Data Streams, In the *Proc. 38th Symp. Interface of Statistics, Computing Science, and Applications (Interface '06)*, 2006.

[24] H. Hotelling. 1931. The Generalization of Student's Ratio, *Annals of Math. Statistics*, Vol. 2, Issue 3, pp. 360-378, 1931.

[25] Hurst, H.E. (1951). "Long-term storage capacity of reservoirs". Transactions of American Society of Civil Engineers. 116: 770.
[26] Hurst, H.E.; Black, R.P.; Simaika, Y.M. (1965). Long-term storage: an experimental study. London: Constable.
[27] SPSS Tutorials: Pearson Correlation. Retrieved 14 May 2017.
[28] Box, G. E. P.; Jenkins, G. M.; Reinsel, G. C. (2008). Time Series Analysis, Forecasting and Control (4th ed.). Hoboken, NJ: Wiley.